\documentclass[11pt]{article}
\usepackage{moriond2000,epsfig}

\def\be{\begin{equation}}
\def\ee{\end{equation}}
\def\bea{\begin{eqnarray}}
\def\eea{\end{eqnarray}}

\begin{document}
\vspace*{4cm}
\title{SEARCH FOR HIGH PROPER MOTION WHITE DWARFS}

\author{ J.T.A. DE JONG, K.H. KUIJKEN \& M.J. NEESER }

\address{Kapteyn Astronomical Institute, Postbus 800, 9700 AV
Groningen, The Netherlands}

\maketitle\abstracts{ 
Recent results of microlensing surveys, show that
10-20 \% of the dark halo mass of the Milky Way consists of compact
objects.  The masses of these compact objects range from 0.1 to 1 solar
mass.  New theoretical cooling models for white dwarfs, and the
detection of faint blue high proper motion objects, imply that ancient
white dwarfs might make up part of this population.\\ 
In this pilot
project, using data from the C and D patches of the ESO Imaging Survey
and data obtained at CTIO, we attempted to find such halo white dwarfs
in the solar neighbourhood.  With a time baseline of approximately one
year between data sets and a limiting I-band magnitude of 23, we can
find these high proper motion objects up to distances of 85 parsecs.  In
the 2.5 deg$^2$ studied so far we have found three high proper motion
candidates. 
}

\section{Introduction: dark matter and white dwarfs}
It is generally accepted that the Milky Way and most external galaxies
contain more mass than observed directly. What makes up this
so-called `dark mass' is, however, not so clear.
This mass is often assumed to be distributed in an isothermal halo,
which means that dark matter would also be present in the solar
neighbourhood. Recent results from galactic microlensing experiments
like MACHO \cite{macho} indicate that 10 to 20 \% of the dark halo
mass of the Milky Way may be in compact objects.

Recent number counts of faint blue objects\cite{mm} in the Hubble Deep Fields
North and South and the detection of high proper motion
objects\cite{ib1,ib2}, indicate the presence of a population of faint
blueish compact objects in the galaxy. Furthermore, new cooling models
for white dwarfs\cite{ha,sj} show that they are much bluer than was
previously believed. The above suggests that the galactic dark halo
might, at least partially, consist of very old white dwarfs.

In the project presented here we try to find faint high proper motion
objects in the solar neighbourhood, which are likely to be dark halo
white dwarfs (see the table on the right).
For this purpose we use the C and D patches of the ESO Imaging Survey
(EIS) Wide\cite{be} and a smaller survey done with the CTIO 4m telescope. 
These data serve as a pilot project for a larger survey for which we
plan to observe the complete C and D patches of the EIS Wide survey with 
a larger temporal baseline.

\section{Data and Method}

The data that are being used for the current project consist of two
overlapping sets.
The first data set is taken from the C and D patches of the EIS Wide
survey\cite{be}, which are 6 $deg^2$ each. These data were obtained
between July 1997 and March 1998 in the I-band with individual exposure
times of 300 seconds. The second data set was obtained with the 4m
telescope at CTIO. This survey covers almost 2 $deg^2$ of both the EIS
C and D patches, making up a total area of nearly 4 $deg^2$. The CTIO
data were taken in the R-band with single pointing exposure times of
2000 seconds. As this second survey was performed in December 1998, we
have a time baseline of approximately one year for the high proper
motion search.

To find high proper motion objects from the data we use a
straightforward method. First, the CTIO data are transformed to the EIS
coordinate system. The EIS pixels are smaller than the CTIO pixels, so
in this way we do not lose information. Following this, objects are matched
within a 2 arcsecond radius. Objects that seem to be offset are selected with
a magnitude dependent cutoff. Finally the resulting outliers are
visually inspected to filter out close objects, cosmic rays, extended
sources, etc. With the time baseline of $\sim$ 1 year we are sensitive
to proper motions of 0.5 to 2.0 arcsecond per year.

\section{What can we find?}

A high proper motion search can be very
efficient in finding halo objects because of their large velocities
with respect to the solar standard of rest. While the sun follows the
galactic rotation, the halo does not. Therefore, halo objects have
typical velocities of 200 km/s.

In the event that we find an object with a high proper motion, how do
we know whether it's part of the dark halo? In table \ref{tab:pops} we
compare the chances of finding objects belonging to the galactic components
present in the solar neighbourhood. Our sensitivity to proper motions
of 0.5 to 2.0 arcseconds, together with the typical velocities of
the objects w.r.t. the sun, gives the distance range where we can find these
objects. With these densities we can calculate the average number of objects
per square degree.
This shows that when we find a high proper motion object, it is most likely
part of the dark halo. According to cooling models by Saumon \&
Jacobsen\cite{sj}, white dwarfs as cool as 3000 K have $M_I \sim 16$;
easily detectable in our survey to distances of 85 pc.

\begin{table}[b]
\caption{Calculated estimates of the average number of high proper
motion objects one will find for different galactic populations. For these
calculations, objects of 0.3 $M_\odot$ are assumed.\label{tab:pops}}
\vspace{0.4cm}
\begin{center}
\begin{tabular}{|l|c|c|c|c|}
\hline
Population & v & distances & $\rho$ & objects\\
~ & ($km/s$) & ($pc$) & ($M_\odot pc^{-3}$) & ($deg^{-2}$)\\
\hline
pop.II halo & 200 & 20-85 & $5 \times 10^{-5}$ & {\bf 0.001}\\
thin disk & 30 & 3-13 & 0.05 & {\bf 0.03}\\
thick disk & 60 & 6-25 & 0.003 & {\bf 0.01}\\
dark halo & 200 & 20-85 & 0.01 & {\bf 2}\\
\hline
\end{tabular}
\end{center}
\end{table}

There are, however, some issues that should be mentioned. The two data
sets are in different bands (R and I), and have different noise
properties because of the different exposure times. Other possible
explanations for apparent object motion include close (variable
star) binaries, distant supernovae and even Kuiper belt and Oort cloud
objects. The motions on the sky of the Kuiper and Oort objects are
dominated by parallax, but because our temporal baseline is $\sim$ 1 year
this possibility can not be excluded with our current data.
All of these problems can be solved by doing a third observation run.

\section{Results}

\begin{figure}[p]
\centering
\psfig{figure=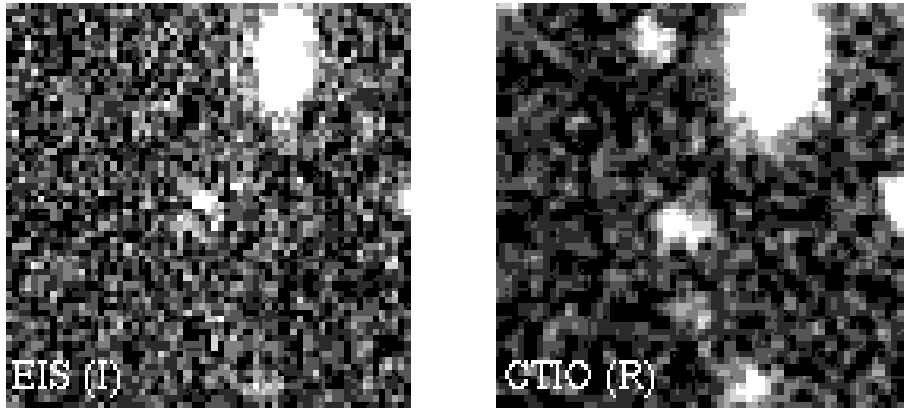,width=12cm}
\caption{Candidate 1. The two images show the same 16x16 arcsec patch of 
the sky (located in the EIS Wide C field).
On the left is the EIS I-band image, on the
right the CTIO 4m R-band image, which was transformed to the EIS coordinate
system. The apparent shift of the image is $\sim$ 1 arcsec.
\label{fig:cand1}}
\end{figure}

\begin{figure}[p]
\centering
\psfig{figure=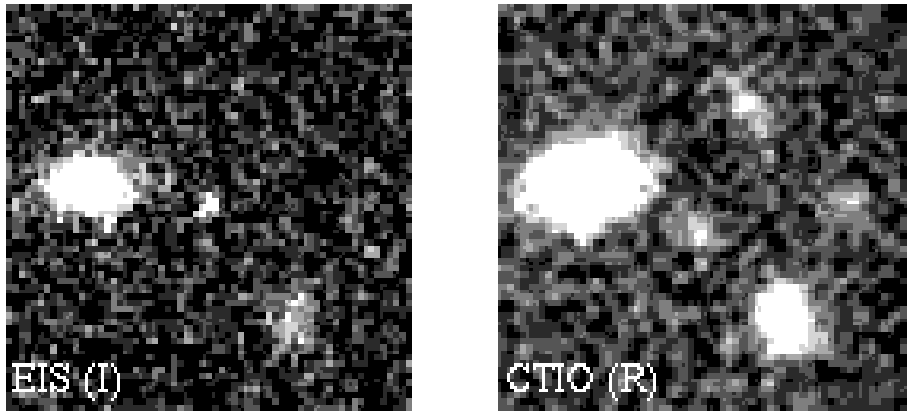,width=12cm}
\caption{Candidate 2. The shift of this candidate is also $\sim$
1 arcsec. This candidate is also located in the EIS Wide C patch and seems
to be moving approximately in the same direction as candidate 1. This
direction is consistent with it being due to the galactic rotation of
the sun.
\label{fig:cand2}}
\end{figure}

\begin{figure}[p]
\centering
\psfig{figure=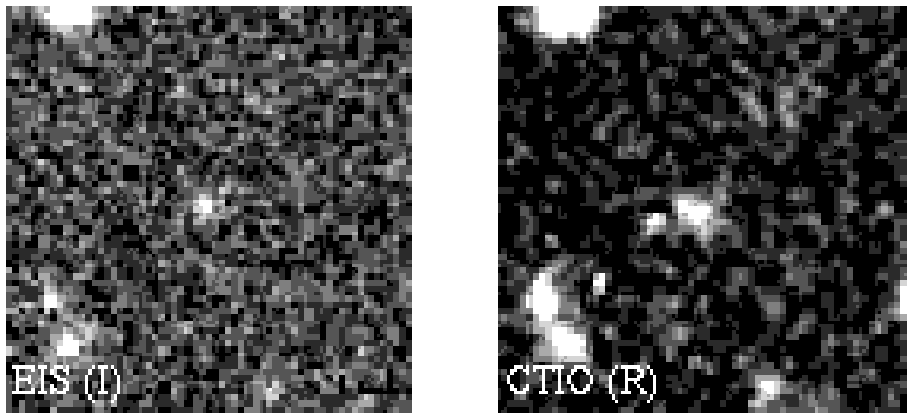,width=12cm}
\caption{Candidate 3. This candidate has an apparent shift of $\sim$
0.5 arcsec and is near our brightness and proper motion detection limits.
The direction of motion of this object is also  consistent with it being 
due to the galactic rotation of the sun. This candidate is located in the 
EIS Wide D patch.
\label{fig:cand3}}
\end{figure}

We have now analyzed 5/8 of the nearly 4 $deg^2$ of data, and 
have found three promising candidates for dark halo objects. These
are shown in the figures \ref{fig:cand1} to \ref{fig:cand3}. The first
two candidates are located in the C patch of the EIS Wide survey while
the third one is in the D patch.

The objects are very faint, which is a further indication that they are not
normal stars. Both candidate 1 and 2 have an apparent I-band magnitude
of 22. The third candidate is even fainter, around $m_I \sim$ 23.
It is also striking that the two C patch objects move
approximately in the same direction. This direction is consistent with
it being due to the galactic rotation of the sun. The same is true for the
motion of the D patch candidate.


\section{Future}

The recently discovered population of faint, high proper motion
objects which may be part of the galactic dark halo needs to be
studied in detail. For this purpose, a large number of objects is
necessary. Therefore, we see the current project as a pilot project
for a larger survey. The most efficient way of increasing the chances
of finding these objects is to increase the field of view. For the
candidates we have so far, further observations need to be done to
confirm the proper motions and to get more color information and/or spectra.

\end{document}